\documentclass[aps,prb,reprint,amsmath,amssymb,aps,superscriptaddress, showpacs]{revtex4-1}

\usepackage{graphicx}% Include figure files
\usepackage{dcolumn}% Align table columns on decimal point
\usepackage{bm}% bold math

\begin{document}

%\preprint{}

\title{Theory of Optical Rectification Effect in Metallic Thin Film with Periodic Modulation}


\author{Hiroyuki Kurosawa}
\email[Present address: ]{kurosawa@es.hokudai.ac.jp}
\author{Seigo Ohno}

%\thanks{}
\affiliation{Department of Physics, Graduate School of Science, Tohoku University, Sendai 980-8578, Japan}

\author{Kazuyuki Nakayama}
\affiliation{Department of Physics, Graduate School of Science, Tohoku University, Sendai 980-8578, Japan}
\affiliation{Center for the Advancement of Higher Education, Tohoku University,  Sendai 980-8576, Japan}


\date{\today}

\begin{abstract}
We conducted theoretical and numerical investigations of the optical rectification (OR) effect in metallic structures with periodic modulation. A new formulation of the OR effect is presented, and the mechanism by which the OR effect is generated, which has been a controversial issue in previous studies, is clarified. We reveal that the OR effect is strongly enhanced by a combination of spatial variation of the metallic structure and local electric field enhancement. Our theory was numerically evaluated and agreed fairly well with experiment.
\end{abstract}

%\keywords{}

\maketitle

\section{Introduction}
Symmetry plays an important role in physics, including nonlinear optics, where it serves as the key feature. Among various kinds of symmetries, space inversion symmetry (SIS) plays the essential role in second-order nonlinear optical processes such as second-harmonic generation (SHG). Second-order nonlinear polarization can be written as
\begin{eqnarray}
P_i^{(2)} =  \chi _{ijk}^{(2)}E_jE_k +Q_{ijkl}E_j\nabla _k E_l, \label{G2ndPol}
\end{eqnarray}
where $\chi  _{ijk}^{(2)}$ and $Q_{ijkl}$ are the second-order nonlinear susceptibility tensors of third and fourth rank, respectively. The first term in Eq. (\ref{G2ndPol}) is called as a dipole term which is associated with the SIS of a material structure. The second term in Eq. (\ref{G2ndPol}) is called as a quadrupole term  and remains significant value even when the material has SIS at an atomic or molecular scale. In mesoscopic artificial structures, such as photonic crystals, plasmonic crystals,  or metamaterials, the spatial profile of electromagnetic fields varies rapidly. Therefore, the quadrupole term can significantly contribute to the nonlinear optical response in those nanostructures, even if they consist of materials with centrosymmetry. SHG, owing to the quadrupole term, has been widely studied in various kinds types of structures such as photonic crystals, metamaterials, and nanoparticles. \cite{PhysRevLett.91.253901, PhysRevB.78.195416, Klein28072006, Martti1038} Symmetries such as chirality and rotational symmetry, as well as SIS, has an intriguing role on the polarization state of SHG irradiated from nano-structures.\cite{PhysRevLett.104.127401, PhysRevLett.112.135502} Terahertz (THz) wave generation due to the difference frequency generation (DFG) has also been reported by Welsh {\itshape et al}. in a metallic grating slab with surface plasmon polariton resonance.\cite{PhysRevLett.98.026803, OptExpress.17.2470} Recently, optical rectification (OR) effects in metallic nano-structures \cite{OptExpress.16.8236, OptLetters.37.2793, PhysRevB.84.035447, 1367-2630-15-11-113061} have received a considerable attention because of their ultra-fast optical response and their potential for the THz optics, which have been theoretically and numerically investigated by Durach {\itshape et al}.\cite{PhysRevLett.103.186801} There are two remarkable theoretical approaches to the formulation of the OR effect in a metallic structure. The first approach, which was developed by Goff and Schaich \cite{PhysRevB.56.15421, PhysRevB.61.10471}, is based on the Euler's equation of fluid motion and corresponds to the viewpoint of a closed circuit. Recently, this theory was applied to the calculation of OR voltage induced in a Au grating slab and reproduced the main feature of experiment. \cite{:/content/aip/journal/apl/98/19/10.1063/1.3590200} The second approach is based on the Lorentz force on bound current and charge and was developed by Durach {\itshape et al}.; this approach corresponds to the viewpoint of an open circuit. Theoretical models that are quite similar to the latter can be found in literature. \cite{PhysRevLett.103.103906, OptExpress.20.1561} Although, owing to the complex formulation of the former theory, it is not easy to obtain an intuitive understanding of the OR voltage, nonetheless, it is applicable to a wide range of problems. This is because the theory is based on electric current, hence the OR voltage can be easily calculated from Ohm's law. On the other hand, although the latter theory is attractive because of its simple formulation and gives an intuitive understanding of the voltage, it is unclear how to convert force to voltage when a metallic structure has a complicated spatial profile. Therefore, the application of the latter has so far been limited to simple structures such as nanowaires and plane thin films.

In this paper, we derive a new formulation of the OR effect in a metallic thin film, with a periodic modulation based on the Euler's equation of fluid motion and on the theory developed by Goff and Schaich, and we show that this theory can be simplified until it is no more complicated than the theory developed by Durach {\itshape et al}. We also present a comparison of the theory with experiment.

This paper is organized as follows. In Section I\hspace{-.1em}I, we present a detailed description of the theory. We provide some analytical formulae for the OR voltage in metallic nano-structures with a periodic modulation, and describe the mechanism of the OR voltage generation. In section I\hspace{-.1em}I\hspace{-.1em}I, we present numerical results for the OR voltage in a metallic nano-structure and a metallic plane film. From these numerical examples, the validity of our theory is clearly confirmed. We conclude the paper in Section I\hspace{-.1em}V with a summary of our results and conclusions. Calculational details of the OR voltage are presented in appendices A and B. The difference between the OR voltage and polarization is described in Appendix C.

\section{Formalism}

 Let us start with Euler's equation of motion for a point charge, $q$, subject to the Lorentz force:
\begin{eqnarray}
\frac{\partial \bm{v}}{\partial t}+ (\bm{v}\cdot \nabla)\bm{v}+\gamma \bm{v} = \frac{q}{m}\left [ \bm{E}+\frac{1}{c}\bm{v}\times \bm{B} \right ], \label{EqnMotion}
\end{eqnarray}
where $\bm{v}, \gamma, m, \bm{E}$, and $\bm{B}$ are the velocity of the charge, phenomenological damping factor, mass of the charge, electric field, and magnetic field, respectively. For simplicity, in Eq. (\ref{EqnMotion}), we have not included the pressure, ($\nabla p; p=\zeta n^{5/3}$), where $\zeta$ and $n$ the constant and density of a free carrier, respectively. Such a simplification can be found in some literature related to SHG, \cite{PhysRevB.79.235109, Shen_PrinciplesNonolinearOptics} while the pressure term is considered in the hydrodynamic theory of photon drag effect developed by Goff and Schaich. Multiplying both sides of Eq. (\ref{EqnMotion}) by the charge density $\rho$ and combining the current density continuity equation, $\partial \rho /\partial t+ \nabla \cdot \bm{j} = 0$ (with $ \bm{j}=\rho \bm{v})$, we obtain the following nonlinear equation for current density $\bm{j}$.
\begin{eqnarray}
 \frac{\partial \bm{j}}{\partial t} +(\nabla \cdot \bm{j})\bm{v} + (\bm{j}\cdot \nabla)\bm{v}+\gamma \bm{j} \label{NLEqn_j} = \frac{q}{m}\left [\rho \bm{E}+\frac{1}{c}\bm{j}\times \bm{B} \right ]. 
\end{eqnarray}
Here, we expand a quantity $\bm{A}$ to the form (where $\bm{A}$ could be $\bm{v},\bm{j},\bm{E},\bm{B}$ or $\rho$): $\bm{A} = \bm{A}^{(0)} +  \bm{A}^{(1)} + \bm{A}^{(2)} + \cdots $, where the superscripts $(0),(1)$ and $(2)$ denote the zeroth-, first-, and second-order responses of the labeled quantity, respectively. In the equilibrium state,  $\bm{v}^{(0)}$, and hence $\bm{j}^{(0)}$ can be set to zero vector. We consider the case where external DC electric and magnetic fields are absent, i.e., $\bm{E}^{(0)} = \bm{B}^{(0)} = \bm{0}$. The first-order of Eq. (\ref{NLEqn_j}) gives us the Drude-type optical response:
\begin{eqnarray}
 \frac{\partial \tilde{\bm{j}}^{(1)}}{\partial t} +\gamma \tilde{\bm{j}}^{(1)} = \frac{q}{m}\rho ^{(0)} \tilde{\bm{E}}^{(1)},
\end{eqnarray} 
from which we can obtain the first order velocity field, $\tilde{\bm{v}}^{(1)} = \tilde{\bm{j}}^{(1)}/\rho ^{(0)} = \left ( iq/m(\omega +i\gamma) \right )\tilde{\bm{E}}^{(1)}$.
The second-order current density is given by
\begin{eqnarray}
&&\frac{\partial \bm{j}^{(2)}}{\partial t} +(\nabla \cdot \bm{j}^{(1)})\bm{v}^{(1)} + (\bm{j}^{(1)}\cdot \nabla)\bm{v}^{(1)}+\gamma \bm{j}^{(2)}\nonumber \\
&&= \frac{q}{m}\left [\rho ^{(1)} \bm{E}^{(1)}+\frac{1}{c}\bm{j}^{(1)}\times \bm{B}^{(1)} \right ]+\frac{q}{m}\rho ^{(0)}\bm{E}^{(2)} \label{2LEqn_j}, 
\end{eqnarray}
where we have assumed that first-order $\bm{E}$ and $\bm{B}$ are monochromatic fields oscillating with frequency $\omega$. Taking the time average of Eq. (\ref{2LEqn_j}) and using it to derive the steady state current, we obtain
the DC component ($\omega - \omega$) of the second-order current density $\bm{j}_{\mathrm{DC}}^{(2)}$:
\begin{widetext}
\begin{eqnarray}
\bm{j}_{\mathrm{DC}}^{(2)} &=& \frac{q}{m\gamma }\frac{1}{2} \mathrm{Re}\left \{ \tilde{\rho }^{(1)}\tilde{ \bm{E}}^{(1)*}+\frac{1}{c}\tilde{\bm{j}}^{(1)}\times \tilde{\bm{B}}^{(1)*} \right \}\label{TA2LEqn_j}- \frac{1}{\gamma }\frac{1}{2}\mathrm{Re}\left \{ (\nabla \cdot \tilde{\bm{j}}^{(1)})\tilde{\bm{v}}^{(1)*} + (\tilde{\bm{j}}^{(1)}\cdot \nabla)\tilde{\bm{v}}^{(1)*}  \right \}+ \sigma ^{(\mathrm{e})}\bm{E}^{(2)}_{\mathrm{DC}} \\
&\equiv& \bm{i}_{\mathrm{DC}}^{(2)} + \sigma ^{(\mathrm{e})}\bm{E}^{(2)}_{\mathrm{DC}},
\end{eqnarray}
\end{widetext}
where $\mathrm{Re} $ and the symbol $*$ denote the real part and complex conjugate of a quantity, respectively. We have also used the relationship $\rho ^{(0)}=qn^{(0)}$ and the electric conductivity defined by the relationship:  $ \sigma ^{(\mathrm{e})} = n^{(0)}q^2/(m\gamma )$ (with $n^{(0)}$ being the density of electric charge) in the derivation of the equation above. We are now in a position to derive voltage in the $x$ direction. From the viewpoint of a closed circuit, the second-order DC electric field is generated by the whole system, and the DC electric field is written as a second-order DC potential $\phi _{\mathrm{DC}}^{(2)}$: $\bm{E}_{\mathrm{DC}}^{(2)} = -\nabla \phi _{\mathrm{DC}}^{(2)}$, which does not have a substantial effect in a closed circuit as the closed line integral $\oint \bm{E}_{\mathrm{DC}}^{(2)} \cdot d\bm{r}= 0$.
Therefore, the current density, which is responsible for the voltage across the structure, is obtained as $\bm{j}_{\mathrm{DC}}^{(2)} = \bm{i}_{\mathrm{DC}}^{(2)}$.

 Ohm's law gives us the voltage induced in the interval $dx$: $dV_{x,\mathrm{DC}}^{(2)}= ( \rho ^{(\mathrm{e})}dx/S_x )\int dS_x i_{x,\mathrm{DC}}^{(2)}$, where $\rho ^{(e)}$ and $S_x$ are the resistivity and cross section of the conductive material at $x$, respectively. Considering that $\rho  ^{(\mathrm{e})}$ is given by the relationship: $\rho  ^{(\mathrm{e})} = 1/\sigma ^{(\mathrm{e})} = m\gamma/ (n^{(0)}q^2)$,  we can obtain the voltage  induced in the $x$ direction
\begin{eqnarray}
V_{x,\mathrm{DC}}^{(2)} = \frac{m\gamma }{n^{(0)}q^2}\int dx \frac{1}{S_x}\int dS_x i_{x,\mathrm{DC}}^{(2)}. \label{FormulaV}
\end{eqnarray}
Hereafter we will omit the superscript $^{(1)}$ of the electric and magnetic fields for simplicity. The terms in the first curly braces $\{ \}$ of $\bm{i}_{\mathrm{DC}}^{(2)}$ are equivalent to the Lorentz force on a bound charge and current, which can be written as
\begin{eqnarray}
(\tilde{\bm{P}}\cdot \nabla )\tilde{\bm{E}}^* + \frac{1}{c}\frac{\partial \tilde{\bm{P}}}{\partial t}\times \tilde{\bm{B}}^* -\nabla \cdot \left (\tilde{\bm{P}}\otimes \tilde{\bm{E}}^* \right ), \label{LorentzF}
\end{eqnarray}
where we have used the relationship: $\tilde{\rho }^{(1)}=-\nabla \cdot \tilde{\bm{P}},\  \tilde{\bm{j}}^{(1)}=\partial \tilde{\bm{P}}/\partial t$ and the vector identity: $\nabla \cdot \left (\tilde{\bm{P}}\otimes \tilde{\bm{E}}^* \right ) = (\nabla \cdot \tilde{\bm{P}})\tilde{\bm{E}}^* + (\tilde{\bm{P}}\cdot \nabla )\tilde{\bm{E}}^*$.
In the vector identity, the symbol $\otimes$ denotes the outer product. The time averaged value of the first and  second terms in (\ref{LorentzF})  can be expressed in the form\cite{PhysRevA.8.14, PhysRevA.58.4279, PhysRevLett.102.113602}:
\begin{eqnarray}
\frac{\alpha _R(\omega)}{4}\nabla  |\tilde{E}|^2 + \frac{\alpha _I(\omega)}{2}\mathrm{Im}\left \{ \tilde{E}_j^*\nabla \tilde{E}_j \right \}, \label{2ndForce}
\end{eqnarray}
where  $\mathrm{Im}$ denotes the imaginary part of the quantity, and $|\tilde{E}|^2, \tilde{P}_j $ and $ \tilde{E}_j,$ are the intensity of the electric field, the complex amplitude of polarization and  electric field, respectively. $\alpha (\omega) = \alpha _R(\omega)+i\alpha _I(\omega)$ is the complex polarizability of the material and $j$ denotes the spatial coordinates: $j=x,y,z$. Summation over $j$ is done according to the Einstein summation convention. The first and second terms of Eq. (\ref{2ndForce}) are called as the gradient force and scattering force, respectively. Now, we will focus on the third and the fourth terms in $\bm{i}_{\mathrm{DC}}^{(2)}$. Substituting $\tilde{\bm{j}}^{(1)}=-i\omega \tilde{\bm{P}}$ and $\tilde{\bm{v}}^{(1)} = \left ( iq/m(\omega +i\gamma) \right )\tilde{\bm{E}}^{(1)}$, we can modify and approximate the quantity $(m\gamma/q)\bm{i}_{\mathrm{DC}}^{(2)}$ of the third and fourth term into the form: 
\begin{eqnarray}
\frac{1}{2}\mathrm{Re}\left \{\frac{1}{1-i\gamma/\omega} \nabla \cdot \left ( \tilde{\bm{P}}\otimes \tilde{\bm{E}}^* \right ) \right \}. \label{iDC2_34}
\end{eqnarray}
Summation of (\ref{iDC2_34}) and the last term in (\ref{LorentzF}) yields the result:
\begin{eqnarray}
\frac{1}{2}\mathrm{Re}\left \{ \frac{\nabla \cdot \left ( \tilde{\bm{P}}\otimes \tilde{\bm{E}}^* \right )}{(i\gamma /\omega)^{-1} -1} \right \} \simeq -\frac{\gamma }{\omega} \frac{1}{2}\mathrm{Im} \left \{  \nabla \cdot \left ( \tilde{\bm{P}}\otimes \tilde{\bm{E}}^* \right )  \right \}. \nonumber\\
\end{eqnarray} 
This approximation is valid where $\gamma / \omega $ is sufficiently small compared with unity. In the near infrared region, the ratio $\gamma / \omega $ is approximately $0.05$ \cite{Rakic:98} and the approximation is valid such that.  Equation (\ref{FormulaV}) may be explicitly described as

\begin{widetext}
\begin{eqnarray}
V_{x,\mathrm{DC}}^{(2)} \simeq \frac{1}{n^{(0)}q}\int \frac{dx}{S_x}\int dS_x \bigg [ \frac{\alpha _R(\omega )}{4}\nabla |\tilde{E}|^2 + \frac{\alpha _I(\omega )}{2}\mathrm{Im}\left \{ \tilde{E}_j^*\nabla \tilde{E}_j \right \} -\frac{\gamma }{\omega} \frac{1}{2}\mathrm{Im} \left \{  \nabla \cdot \left ( \tilde{\bm{P}}\otimes \tilde{\bm{E}}^* \right )  \right \}  \bigg ]_x .\label{Formula_mainV2}
\end{eqnarray}
\end{widetext}

This is the main result of this paper and the formula for the voltage generated by the OR effect in a metallic structure. As evident from the derivation, voltage in terms of electric current density can be expressed in terms of the Lorentz force, and the relationship between voltage and the force density is clarified. Note that the voltage was evaluated using the volume integral of the force on the basis of the Lorentz force in previous studies. We found that the volume integral of the force is not always valid for the evaluation of voltage and is limited to a structure with a constant cross section such as a nanowire and a plane thin film. Furthermore, we have explicitly derived the third term in Eq. (\ref{Formula_mainV2}) for the first time. This term is of importance for the OR effect in a metallic film, and we will discuss its physical interpretation later. 

\subsection{Periodically corrugated metallic nano-structure}

\begin{figure}
 \centering
 \includegraphics[width=8.5cm]{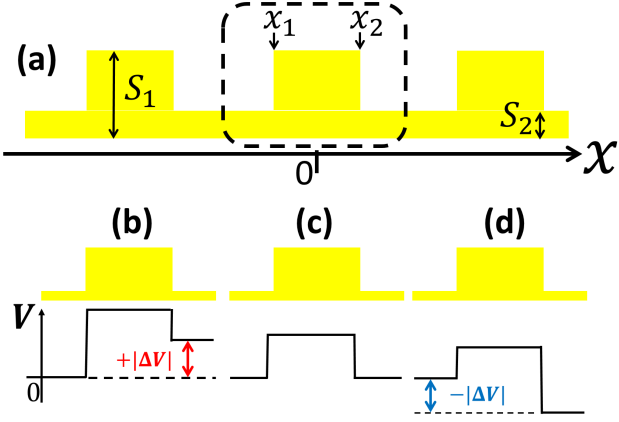}
 \caption{(Color online) (a): Schematic of the periodic metallic structure.  The unit cell is surrounded by the dashed line.  The spatial profile of the OR voltage across the metallic grating slab in case of positive (b), zero (c), and negative (d) voltage.}
 \label{Fig1}
\end{figure}

We first consider the OR voltage induced in a periodically corrugated metallic nano-structure. In this subsection, we ignore the second and third terms in Eq. (\ref{Formula_mainV2}), which are small compared with the first term in Eq. (\ref{Formula_mainV2}) because of the relation $\alpha _I (\omega)=(\gamma /\omega )\alpha _R(\omega)$ for Drude metals. Therefore, we focus on the voltage induced by the first term in Eq. (\ref{Formula_mainV2}) and consider the voltage in the $x$ direction. Hence we take the spatial derivative along that axis: $\nabla \rightarrow  \partial _x$.  In this case, the voltage  due to the gradient force is given by
\begin{eqnarray}
V_{x,\mathrm{DC}}^{(2)}=\frac{1}{n^{(0)}q}\int dx \frac{1}{S_x}\int dS_x \frac{\alpha _R}{4}\frac{\partial }{\partial x} \left | \tilde{E}(\bm{r}) \right |^2. \label{GradV}
\end{eqnarray}
Using the partial integration method, we can modify Eq. (\ref{GradV}) to the form:
\begin{eqnarray}
V_{x,\mathrm{DC}}^{(2)}&=&\frac{1}{n^{(0)}q}\int dx \bigg \{ -\frac{\partial}{\partial x}\left (\frac{\left | \alpha _R\right |}{4}\left \langle \left | \tilde{E}(\bm{r}) \right |^2 \right \rangle _S \right ) \nonumber \\&&\hspace{15mm}+ \frac{\alpha _R}{4}\left \langle \left | \tilde{E}(\bm{r}) \right |^2 \right \rangle _S \frac{1}{S_x}\frac{\partial S_x}{\partial x} \bigg \} \label{GradV2_mainText},
\end{eqnarray}
where we have introduced the notation $\left \langle A(x) \right \rangle _S = (1/S_x(x))\int dS_x A(x)$, which corresponds to the surface average of a function $A(x)$. We have also replaced $\alpha _R$ with $-|\alpha _R|$ in the first term of Eq. (\ref{GradV2_mainText}), which is valid for noble metals in the optical region because of the negative value of the real part of the polarizability in that region. From the viewpoint of kinematics, the first term in Eq. (\ref{GradV2_mainText}) corresponds to the force due to potential ($-\nabla U,$ where $U = |\alpha _R| \left \langle |\tilde{E}|^2\right \rangle _S/4$) and the potential is periodic due to the periodic boundary condition. Moreover,  the rotation of the gradient force is exactly zero:
\begin{eqnarray}
 \nabla \times \nabla \left ( \frac{\left | \alpha _R \right |}{4}\left \langle \left | \tilde{E}(\bm{r}) \right |^2 \right \rangle _S \right )= 0,
\end{eqnarray}
which indicates that the gradient force is conservative force. Therefore, the first term in Eq. (\ref{GradV2_mainText}) disappears when integrated along the $x$ direction due to the periodic boundary condition. Note that we have assumed that the potential $U$ is smooth along the $x$ axis, which is not always valid. For example, the potential $U$ is not smooth in some spiky structures such as sharply brazed grating. In that case, the derivative of $U$ yields delta function owing to the non-smooth function and hence the first term in Eq. (\ref{GradV2_mainText}) remains significant value. In this paper, we deal with structures with smooth spatial profile of electric field intensity. Thus, the second term is responsible for the generation of the voltage across the structure. As long as we focus on not the local voltage but the voltage across the structure, we do not need to pay attention to the contribution from the first term in Eq. (\ref{GradV2_mainText}). Then we finally reach the most simplified expression for the voltage across the structure in the $x$ direction:
\begin{eqnarray}
V_{x,\mathrm{DC}}^{(2)}=\frac{1}{n^{(0)}q}\int dx \left \{ \frac{\alpha _R}{4}\left \langle \left | \tilde{E}(\bm{r}) \right |^2 \right \rangle _S \frac{1}{S_x}\frac{\partial S_x}{\partial x} \right \}, \label{FormulaC}
\end{eqnarray}
which indicates that the voltage across the structure is generated where the cross section of the structure varies spatially. In previous studies, the enhancement and confinement of the electric field were focused in order to obtain higher voltage. Our study reveals that not only higher intensity of electric field but also strong relative variation of the cross section of the structure are essential for the higher OR voltage.  To understand the generation of the OR voltage, let us consider the metallic Lamellar grating shown in Fig. \ref{Fig1}(a). The grating has spatial variation at the positions of $x_1$ and $x_2$  in its unit cell and the derivative of its spatial profile at $x_1$ has the opposite sign of that of $x_2$. When the local electric field intensity at $x_1$ is different from that of $x_2$, the symmetry of the local voltage is broken and uneven, which results in a non-zero voltage across the structure (Figs. \ref{Fig1}(b) and (d)). Note that voltage is not induced across the structure when the local electric field at $x_1$ and $x_2$ is even (Fig. \ref{Fig1}(c)).

\begin{figure}
 \centering
 \includegraphics[width=8.5cm]{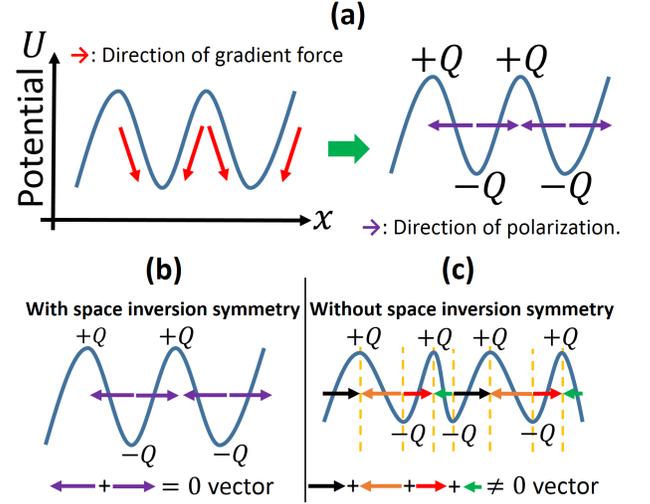}
 \caption{(Color online) (a): Free carriers subject to a periodic potential (shown by blue curve) induce polarization in a structure. Conceptual schematic of total polarization induced by the gradient force in a structure with space inversion symmetry (SIS) (b) and without SIS (c). }
 \label{Fig2}
\end{figure}

In the end of this subsection, we briefly discuss the role of the gradient term which was ignored in the evaluation of the OR voltage. Even though the force due to the potential $(-\nabla U)$ does not contribute to the OR voltage, it can induce local polarization in a structure as is schematically shown in Fig. \ref{Fig2}(a).  In general, the volume integral of the  gradient force remains significant value:
\begin{eqnarray}
\int dV \frac{\alpha _R}{4}\nabla \left | \tilde{E}(\bm{r}) \right |^2 = \oint d\bm{S}  \frac{\alpha _R}{4}\left | \tilde{E}(\bm{r}) \right |^2, \label{V2S}
\end{eqnarray}
where we have used Gauss' theorem to convert the volume integral to the surface one. Equation (\ref{V2S}) corresponds to the driving force which induces mesoscopic polarization in metallic structures.  When the SIS is not broken at the scale comparable with the period of the structure, mesocopic polarization is not induced in the structure, even though local polarization is induced (shown in Fig. \ref{Fig2}(b)). On the other hand, mesoscopic polarization remains significant value when the SIS is broken at the mesoscopic scale (shown in Fig. \ref{Fig2}(c)). In that case, Eq. (\ref{V2S}) gives significant value. Thus, it is concluded that the gradient force, which is conservative force, can induce mesoscopic polarization in a structure but it does not contribute to the OR voltage across the structures. Displacement of an electric charge from the equilibrium state, $ \bm{d}$, is proportional to the gradient force. Therefore, the mesoscopic polarization induced in a structure, $\bm{P}$, is related to the local polarization, $q\bm{d}$, and the gradient force as follows:
\begin{eqnarray}
\bm{P} &=& \int dV q\bm{d}\\
          &\propto &  \int dV q\frac{\alpha _R}{4}\nabla \left | \tilde{E}(\bm{r}) \right |^2 =  q\frac{\alpha _R}{4} \oint d\bm{S} \left | \tilde{E}(\bm{r}) \right |^2. \label{Pol}
\end{eqnarray}

We present the relationship between the OR voltage $V_{x,\mathrm{DC}}^{(2)}$ and DC polarization $\bm{P}$ in appendix C.

\subsection{Metallic plane film}

From Eq. (\ref{FormulaC}), it is clear that the OR voltage induced by the gradient force disappears in structures with a constant cross section, such as metallic nanowires and plane films with periodicity. Namely, the OR voltage in these structures is generated by the second and third terms in Eq. (\ref{Formula_mainV2}). Therefore, we can obtain the voltage induced in a structure with a constant cross section $S_x(x) = S$ as follows:
\begin{eqnarray}
V_{x,\mathrm{DC}}^{(2)}\simeq && \frac{1}{n^{(0)}q}\frac{1}{S} \int dV  \bigg [ \frac{\alpha _I(\omega )}{2} \mathrm{Im}\left \{ \tilde{E}_j^*\nabla \tilde{E}_j \right \}\notag \\
&&\hspace{10mm} -\frac{\gamma }{\omega} \frac{1}{2}\mathrm{Im} \left \{ \oint dS \left ( \tilde{\bm{P}}\cdot \hat{\bm{n}} \right ) \tilde{\bm{E}}^* \right \}  \bigg ]_x,\label{FormulaV_unformS}
\end{eqnarray}
where we have applied Gauss' law to convert the volume integral to a surface one in the second term in Eq.( \ref{FormulaV_unformS}) and $\hat{\bm{n}}$ is the unit vector directed outwards from the structure\rq{}s surface. This is the formula for the voltage in a structure with a constant cross section. Taking into account that the surface charge density, $\tilde{\sigma} _s$, is given by the relationship $\tilde{\sigma} _s= \tilde{\bm{P}}\cdot \hat{\bm{n}}$, we can conclude that the second term in Eq. (\ref{FormulaV_unformS}) corresponds to the OR voltage due to the surface charge, which was not formulated in previous studies. We also found that OR effect of the surface charge is due to  dissipative processes, because it is expressed by the imaginary part of the force density and is proportional to the damping factor $\gamma$.

\section{Comparison of the theory with experiment}

\subsection{Periodically corrugated metallic nano-structure}

\begin{figure}
 \centering
 \includegraphics[width=8.5cm]{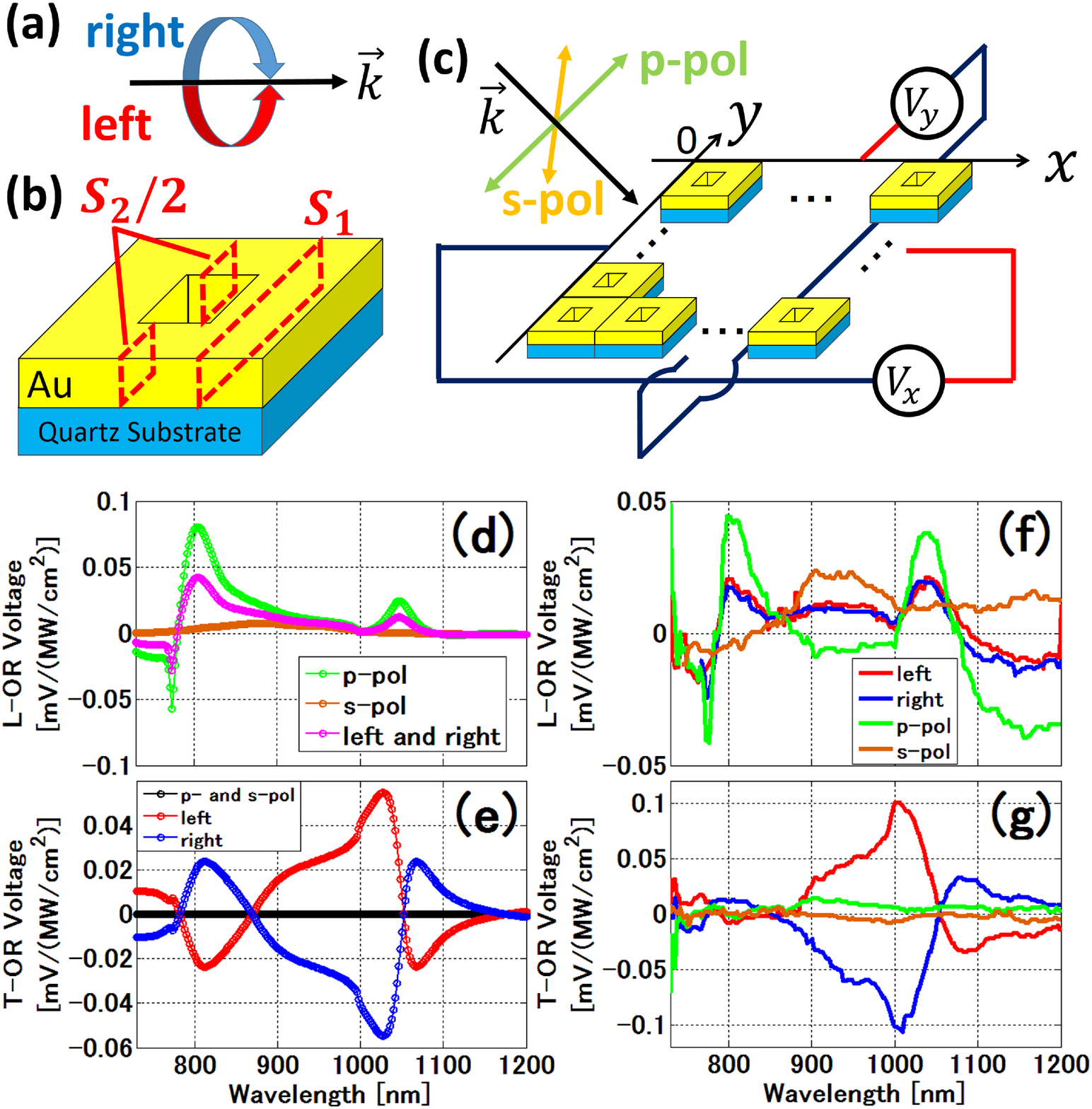}
 \caption{(Color online) (a): Definition of left (red) and right (blue) circular polarization. (b): Schematic of unit cell of the metallic photonic crystal slab. (c): Configuration of longitudinal and transverse optical rectification voltage. (d), (e): Calculation result of the longitudinal (d) and the transverse (e) OR voltage spectra. (f), (g): Experimental data on the longitudinal (f) and transverse (g) OR voltage spectra. The data was taken from the literature.\cite{PhysRevLett.103.103906}}
 \label{Fig3}
\end{figure}

To confirm the validity of our theory, let us apply Eq. (\ref{FormulaC}) to a numerical calculation of the voltage induced in a 2D metallic grating slab with an air-hole array, shown in Fig. \ref{Fig3}(c). The 2D grating has a  period of 500 nm in the $x$ and $y$ directions. A 250 nm-square air-hole array is perforated in a 40 nm-thick Au film. We performed numerical calculation in the case of incident angle 33$^\circ$. Sample and experimental conditions are the same as those in the literature.\cite{PhysRevLett.103.103906} Local electromagnetic fields are calculated by the scattering matrix method.\cite{PhysRevB.66.045102, JOSAA.13.1024, JOSAA.14.2758} In the grating structure, the cross section around $x=x_1$ in the unit cell can be modeled as  
\begin{eqnarray}
S_x(x)\bigg |_{x\sim x_1} = \frac{\left( S_1-S_2\right )}{\pi}\tan ^{-1}\frac{x-x_1}{\epsilon} + \frac{S_1+S_2}{2},\nonumber \\
\end{eqnarray}
where $\epsilon$ is the infinitesimal small number, and $S_1$ and $S_2$ are the largest and smallest cross sections of the 2D metallic grating slab ($S_1 = 500$ nm  $\times 40$ nm, $S_2=250 $ nm $\times 40$ nm), respectively. The cross section function, $S_x(x)$, is similar to the step function  near $x=x_1$, and also tends to $S_2$ and $S_1$ in the limit that $x \rightarrow \mp \infty$, respectively. The derivative of $S_x(x)$ is analytically calculated to be
\begin{eqnarray}
\frac{\partial S_x(x)}{\partial x} &=& \frac{S_1 - S_2}{\pi} \frac{\epsilon }{(x-x_1)^2+\epsilon ^2} \\
&\rightarrow& \left(S_1-S_2\right )\delta (x-x_1),
\end{eqnarray}
 where we have took the limit that $\epsilon \rightarrow 0$ and used the relationship $\lim _{\epsilon \rightarrow 0}\mathrm{Im}\left( x+i\epsilon \right)^{-1}= -\pi \delta (x)$, and $\delta (x)$ is the Dirac delta function. The cross section function around $x_2$ and its derivative can be modeled and calculated as the same manner. Thus the derivative of the cross section function in the unit cell is written as
\begin{eqnarray}
\frac{\partial S_x(x)}{\partial x} = \left(S_1-S_2 \right) \left ( \delta (x-x_1) - \delta (x-x_2) \right ). \label{DerivativeSx}
 \end{eqnarray}
Namely, the OR voltage induced in the Lamellar metallic grating has impulsive contributions at $x_1$ and $x_2$.  After substituting Eq. (\ref{DerivativeSx}) into Eq. (\ref{FormulaC}), we finally obtain the voltage induced in the $x$ direction:
 \begin{widetext}
 \begin{eqnarray}
V_{x,\mathrm{DC}}^{(2)}=\frac{1}{n^{(0)}q}\frac{\alpha _R}{4}\left( 1 -\frac{S_2}{S_1}\right )\left \{ \left \langle \left | \tilde{E}(x_1) \right |^2 \right \rangle _S - \left \langle \left | \tilde{E}(x_2) \right |^2 \right \rangle _S  \right \}. \label{Formula_D}
 \end{eqnarray}
 \end{widetext}
 
Namely, the OR voltage consists of three factors. The first is the material parameter $(\alpha _R/(4n^{(0)}q))$ which is related to Hall coefficient $1/(n^{(0)}qc)$ of the conductive material. The second is $\left (1-S_2/S_1 \right )$ which corresponds to the strength of the spatial modulation of the structure. The remaining factor is related to the strength and asymmetry of the local electric field intensity. These analysis indicates that high Hall coefficient, strong spatial modulation, and asymmetric high intensity of electric field are essential for the emergence of giant OR voltage. In the previous study, the local enhancement of electric field is focused in order to obtain higher voltage. Our theory reveals the essential role of the spatial modulation of the structure. Note that our theoretical treatment is easily extended to the 2D case by simple substitution from $x$ to $y$. Figures \ref{Fig3}(d) and (e) show the calculated longitudinal (L-) and transverse (T-) OR voltage spectra, respectively. On the other hand, Figs. \ref{Fig3}(f) and (g) show their experimental values, which have been taken from the literature.\cite{PhysRevLett.103.103906} The two peaks observed in the L-OR voltage correspond to surface plasmon polariton resonance folded back by the reciprocal lattice vector of the metallic structure. When surface plasmon polariton is excited, as observed in the experiment, the L-OR voltage is at its peak, whereas the T-OR voltage is dispersive. These characteristic features, as well as the spiky structure in L-OR voltage around the wavelength 780 nm, are clearly reproduced by our new formulation of the OR voltage. As is reported in the literature, the polarity of T-OR voltage flips its sign depending on the sense of circular polarization (red and blue lines in Fig. \ref{Fig3}(g)). It is also reported that L-OR voltage spectra for circular polarization is the same as the average of that of p- and s-polarization (see Fig. \ref{Fig3}(f)). All these features are also reproduced in our numerical calculation. Thus, it is confirmed that our theory elucidates the OR voltage in the metallic nano-structure fairly well. Note that the air-hole array is circular in the experiment, while a square air-hole array is considered, for simplicity,  in the numerical calculation. In spite of the difference of the air-hole shape between calculation and experiment, the OR voltage spectra are well reproduced. This result is attributed to the relative spatial variation term in Eq. (\ref{FormulaC}). Namely, the relative spatial variation of a structure is essential in the OR voltage under the condition that the different structures have a similar electromagnetic response and consist of the same materials.

\begin{figure}
 \centering
 \includegraphics[width=8.5cm]{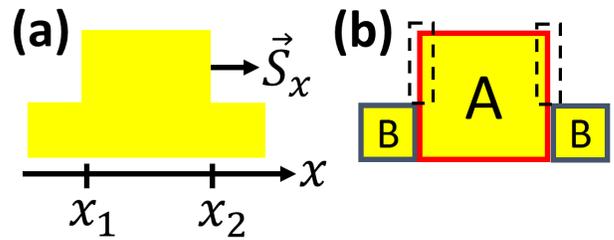}
 \caption{(Color online) (a): Schematic of a Lamellar grating with sharp edge. (b): Definition of parts A and B used in the calculation of the OR voltage. \label{Fig4}}
\end{figure}

So far we considered the case where the spatial profile of electric field intensity was smooth function of $x$ and $y$.  As an ideal case, we can consider q sharply edged rectangular Lamellar grating shown in Fig. \ref{Fig4}(a), where the first term in Eq. (\ref{GradV2_mainText}) cannot be ignored because of the discontinuity of the derivative.  In this case, we first divide the structure into two parts (A and B in Fig. \ref{Fig4}(b)) and calculate the OR voltage induced in each part. Then the total OR voltage is obtained by summing up the OR voltage induced in parts A and B. The calculational details of the OR voltage in this method is described in Appendices A and B. 

\subsection{Metallic plane film}

\begin{figure}
 \centering
 \includegraphics[width=8.5cm]{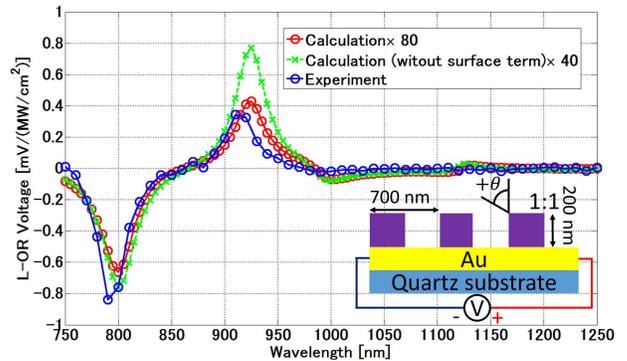}
 \caption{(Color online) Calculated (red) and measured (blue) OR voltage spectra of the metal-dielectric grating shown in the inset. The green line shows the calculation result without the surface term. The experimental data was taken from the literature.\cite{OptExpress.20.1561} \label{Fig5}}
\end{figure}

Finally, let us apply Eq. (\ref{FormulaV_unformS}) to the structure with a constant cross section that is shown in the inset of Fig. \ref{Fig5}, whose OR response was investigated in the literature.\cite{OptExpress.20.1561} The metallic thin film has a 1D dielectric grating with a period of 700 nm.  We performed numerical calculation in the case of incident angle 6$^\circ$. Sample and experimental conditions are the same as those in the literature.\cite{OptExpress.20.1561}  In this metallic thin film, surface plasomon polaritons (SPPs) can be excited under p-polarized excitation, and the peak and dip features in Fig. \ref{Fig5} correspond to the SPPs resonance. The blue and red lines in Fig. \ref{Fig5} are the measured and calculated spectra, respectively, of the L-OR voltage. Our numerical calculation reproduced the experimental result well, except for the overall amplitude of the voltage. To investigate the importance of the surface term, we show the calculation result of the L-OR voltage without the second term of Eq. (\ref{FormulaV_unformS}) (green line). The calculated OR voltage exhibits a symmetric response with respect to the amplitude, whereas the experimental OR voltage clearly exhibits an asymmetric response. Our numerical calculation that includes the surface term (red line) reproduces the asymmetric structure fairly well and shows that the surface term has a significant contribution to the OR voltage induced in the metallic plane film.

\section{Conclusion}

In conclusion, we theoretically and numerically investigated the OR effect in metallic thin film with periodic modulation. We succeeded in formulating the OR voltage induced in the metallic structure and derived quite simple expression of the OR voltage. The derived formula indicates that the OR voltage across the structure is generated where the cross section of the structure varies spatially in metallic nano-structure and spatial variation of the structure plays the essential role in the OR voltage. When a dielectric nano-structure is on a plane metallic thin film, the OR voltage originates from two contributions: the bulk scattering force and surface one. We, for the first time, formulated the surface OR voltage explicitly and found the significant contribution from the surface term. From the theoretical frame-work perspective, we unified the two previous  theoretical models and showed a simple and exact picture of the OR voltage. We also performed numerical calculation and demonstrated that our theory reproduced the experimental result well, which indicates the validity of our theory. Our findings in this study can lead to design criteria for the generation of giant OR voltages in metallic nanostructures with periodic modulation.

\begin{acknowledgments}
The authors are grateful to Professors Teruya Ishihara and Masayuki Yoshizawa at Tohoku University for critical discussion and insightful suggestions. We are also grateful to Makoto Saito at Tohoku University for his technical support. This work has been partly supported by Grant-in-Aid for Scientific Research on Innovative Areas (No. 22109005, Electromagnetic Metamaterials) from the Ministry of Education, Culture, Sports, Science, and Technology (MEXT), Japan. Hiroyuki Kurosawa is sponsored by the Japan Society for the Promotion of Science (JSPS).
\end{acknowledgments}

\appendix

\section{Calculation of the OR voltage induced in a spiky structure}
In the main text, we have assumed that the surface average of electric field intensity is a smooth function along the $x$ direction. Our assumption is consistent with experimental conditions. As an ideal case, we can consider a structure with sharp edges shown in Fig. \ref{Fig4}(a). Let us consider the voltage across the structure due to the optical rectification effect.

Let us first divide the structure into three parts as shown in Fig. \ref{Fig4}(b). According to Eq. (\ref{GradV}), the voltage across the structure given by the sum of the voltage induced in each part. In the divided part bounded by the blue lines in Fig. \ref{Fig4} (let us label the part as B), the cross section of the structure is constant along the $x$ direction. Therefore, the integral in Eq. (\ref{GradV}) can be reduced to the surface integral as follows:
\begin{eqnarray}
V_{x,\mathrm{DC}}^{(2),B}&=&\frac{1}{n^{(0)}q}\frac{1}{S_2}\int dV \left [ \frac{\alpha _R}{4}\nabla \left | \tilde{E}(x) \right |^2 \right ]_x\nonumber \\
     &=&\frac{1}{n^{(0)}q}\frac{1}{S_2}\left [ \oint dS  \frac{\alpha _R}{4} \left | \tilde{E}(x) \right |^2 \right ]_x. \label{GradV2}
\end{eqnarray}
However, we must pay attention to the derivative when calculating Eq. (\ref{GradV}) in the divided part bounded by the red lines (let us label the part as A). Equation (\ref{GradV}) can also be written as
\begin{eqnarray}
V_{x,\mathrm{DC}}^{(2)}&=&\frac{1}{n^{(0)}q}\frac{1}{S_1}\left [ \int dV \frac{1}{2}\mathrm{Re} \left \{ \left ( \tilde{P}_i\nabla \tilde{E}_j \right ) \right \} \right ]_x, \label{GradV3}
\end{eqnarray}
where $\tilde{P}_i$ is the $i$ th component of complex polarization $\tilde{\bm{P}}$ defined as $\tilde{\bm{P}}=\alpha (\omega) \tilde{\bm{E}}$.  In order to evaluate the voltage induced in the divided parts, we need to take the derivative of $\tilde{E}_x$ and $\tilde{E}_z$ with respect to $x$. However, only the part A has the surfaces perpendicular to the $x$ direction where the $x$ component of electric field is discontinuous (the surface bounded by the dashed lines in Fig. \ref{Fig4}(b)). Note that when the incident light is s-polarized (electric field perpendicular to the incident plane) and the structure is 1D, we have only the s-polarization component of electric field. Hence, we do not need to pay attention to the derivative of electric field. Therefore, we must perform special treatment  at the surface when taking the derivative of $\tilde{E}_x$ with respect to $x$.  We divide the integral (\ref{GradV3}) into three region: one is the bulk region and the others are two surfaces perpendicular to the $x$ direction. Namely, the volume integral inside the bracket $[ \ ]_x$ in Eq. (\ref{GradV3}) is decomposed into three regions as follows: 

\begin{eqnarray}
&&\int dV \frac{1}{2}\mathrm{Re}\left \{ P_j^*\nabla E_j \right \}\nonumber \\
&=& \int dS_x \int _{x_1}^{x_2} dx \frac{1}{2}\mathrm{Re}\left \{ P_j^*\nabla E_j \right \}\nonumber \\
&=& \int dS_x\left \{  \int _{x_1}^{x_1+\epsilon} + \int _{x_1+\epsilon}^{x_2-\epsilon} + \int _{x_2-\epsilon}^{x_2} \right \}dx \frac{1}{2}\mathrm{Re}\left \{ P_j^*\nabla E_j \right \},\nonumber \\
\end{eqnarray}
where $\epsilon$ is the infinitesimal positive number and $S_x$ denotes the surface perpendicular to the $x$ axis. As for the integral of gradient force with respect to $x$ from $x_1+\epsilon$ to $x_2-\epsilon$, it is converted to surface integral following Gauss\rq{s} law:
\begin{eqnarray}
&&\int d\bm{S}_x \int _{x_1+\epsilon}^{x_2-\epsilon}dx \frac{1}{2}\mathrm{Re}\left \{ P_j^*\nabla E_j \right \}\nonumber \\
&&= \int _{\partial S_x}d\bm{S}_x \frac{\alpha _R (\omega)}{4}\left ( \tilde{E}_j \tilde{E}_j^* \right ) + \int dV  \frac{\alpha _I(\omega)}{2}\mathrm{Im}\left \{ E_j^*\nabla E_j\right \},\nonumber \\ \label{BulkForce}
\end{eqnarray}
where surface integral is performed just inside the metal surface. Ignoring the scattering force, we obtain the voltage induced in the bulk region: 
\begin{eqnarray}
V_{x,{\mathrm{DC,Bulk}}}^{(2),A} = \frac{1}{n^{(0)}q}\frac{1}{S_1}\left [ \int _{\partial S_x}d\bm{S}_x \frac{\alpha _R (\omega)}{4}\left | \tilde{E}(x) \right |^2 \right ]_x.\nonumber \\ \label{BulkForceV}
\end{eqnarray}

As for the voltage induced at the surface, let us consider the surface bounded by metal and dielectric. At the surface, $x$ component of electric field is discontinuous. Therefore, the derivative of $E_x$ with respect to $x$ can be written as\cite{0953-4075-39-15-S14}
\begin{eqnarray}
\frac{\partial E_x}{\partial x} = \delta (x)\left ( E_x^M (x_1^+) - E_x^D(x_1^-)  \right ), \label{DeltaFunc}
\end{eqnarray}
where $E_x^M$ and $E_x^D$ are the $x$ components of electric field in the metallic and dielectric region, respectively.  Condition of continuity of $x$ component of the electric displacement allows us to rewrite $E_x^D(x_1^-)$ in terms of $E_x^M(x_1^+)$:
\begin{eqnarray}
E_x^D(x_1^-) = \frac{\varepsilon ^M}{\varepsilon ^D}E_x^M(x_1^+),  \label{Conti-D}
\end{eqnarray}
where $\varepsilon ^D$ and $\varepsilon ^M$ are the permittivities of the dielectric and metallic regions, respectively. Therefore Eq. (\ref{DeltaFunc}) can be rewritten as
\begin{eqnarray}
\frac{\partial E_x}{\partial x} = -\delta (x)\left ( \frac{\varepsilon ^M}{\varepsilon ^D} - 1 \right )E_x^M(x_1^+).
\end{eqnarray}
The integral near the surface is explicitly describe as
\begin{eqnarray}
&&\int dS_x \int _{x_1}^{x_1+\epsilon}dx \frac{1}{2}\mathrm{Re}\left \{ \tilde{P}_j^*\nabla _x \tilde{E}_j \right \}\nonumber \\
&&=\frac{1}{2}\mathrm{Re}\left \{  \int dS_x \int _{x_1}^{x_1+\epsilon}dx \left ( \tilde{P}_x^*\partial _x \tilde{E}x + \tilde{P}_z^*\partial _x \tilde{E}_z \right ) \right \}. \nonumber \\ \label{Surf-Force}
\end{eqnarray}
We first focus on the second term of Eq. (\ref{Surf-Force}). Since $E_z$ is continuous at the surface, we do not need to have special treatment of the derivative of $E_z$ with respect to $x$: 
\begin{eqnarray}
&&\int _{x_1}^{x_1+\epsilon} dx \tilde{P}_z^* (x)\partial _x \tilde{E}_z(x)\nonumber \\
&=& \int _{x_1}^{x_1+\epsilon }dx \frac{\tilde{P}_z^*(x_1)+\tilde{P}_z^*(a+\epsilon)}{2}\frac{\tilde{E}_z(x_1+\epsilon)-\tilde{E}_z(x_1)}{\epsilon}\nonumber \\
&=&  \frac{\tilde{P}_z^*(x_1)+\tilde{P}_z^*(x_1+\epsilon)}{2}\frac{\tilde{E}_z(x_1+\epsilon)-\tilde{E}_z(x_1)}{\epsilon} \epsilon \\
&\rightarrow& 0 (\epsilon  \rightarrow 0),\nonumber 
\end{eqnarray}
where we have replaced $\tilde{P}_z^*$ with its space average. Therefore, the second term of Eq. (\ref{Surf-Force})  gives zero.

Next, we focus on the first term of the right hand side of Eq. (\ref{Surf-Force}), the integral with respect to $x$ is given by
\begin{eqnarray}
&&\int _{x_1}^{x_1+\epsilon }dx \tilde{P}_x^*(x) \partial _x \tilde{E}_x \nonumber \\
&=& \int _{x_1}^{x_1+\epsilon }dx \tilde{P}_x^*(x) \left \{ -\delta (x)\left ( \frac{\varepsilon ^M}{\varepsilon ^D} - 1 \right )E_x^M(x_1^+) \right \}\nonumber \\
&=& -\frac{1}{2}\tilde{P}_x^*(x_1) \left ( \frac{\varepsilon ^M}{\varepsilon ^D} - 1 \right ) E_x^M(x_1^+), \label{EqnAD}
\end{eqnarray}
where we have used the formula:
\begin{eqnarray}
\int _0^{\epsilon }f(x)\delta (x)dx = \frac{1}{2}f(0).
\end{eqnarray}
Macroscopic polarization at the surface located at $x$ is given by space average of polarization of the dielectric and metallic region:
\begin{eqnarray}
\tilde{P}_x(x_1) &=& \frac{\tilde{P}_x^{D}(x_1^-) + \tilde{P}_x^{M}(x_1^+)}{2}\nonumber ,\\
&=&  \frac{\alpha ^{D}E_x^{D}(x_1^-) + \alpha ^{M}\tilde{E}_x^{M}(x_1^+)}{2}\nonumber \\
&=& \frac{\alpha ^{D}\varepsilon ^{M}/\varepsilon ^{D} + \alpha ^{M}}{2}\tilde{E}_x^{M}(x_1^+), \label{EqnPol}
\end{eqnarray}
where we have used Eq. (\ref{Conti-D}) to rewrite $E_x^{D}$ in terms of $E_x^{M}$, and  $\alpha ^D$ and $\alpha ^M$ are the complex polarizabilities of the dielectric and the metal, respectively. Substituting Eq. (\ref{EqnPol}) into Eq. (\ref{EqnAD}), we obtain the result below:
\begin{eqnarray}
&&\int _{x_1}^{x_1+\epsilon }dx \tilde{P}_x^*(x) \partial _x \tilde{E}_x\nonumber \\
&=& -\frac{1}{2}\left ( \frac{\alpha ^{D}\varepsilon ^{M}/\varepsilon ^{D} + \alpha ^{M}}{2} \right )^* \left ( \frac{\varepsilon ^M}{\varepsilon ^D} - 1 \right )  | E_x^M(x_1^+) |^2 .\nonumber \\
\end{eqnarray}
Taking surface integral of the equation above, we finally obtain the OR voltage at the surface as
\begin{widetext}
\begin{eqnarray}
\int dS_x \int _{x_1}^{x_1+\epsilon}dx \frac{1}{2}\mathrm{Re}\left \{ \tilde{P}_x^*\nabla _x \tilde{E}_x \right \} = \int d\bm{S}_x \frac{1}{8} \mathrm{Re} \left \{ \left ( \alpha ^{D}\frac{\varepsilon ^{M}}{\varepsilon ^{D}} + \alpha ^{M} \right )^* \left ( \frac{\varepsilon ^M}{\varepsilon ^D} - 1 \right ) \right \}  | E_x^M(x_1^+) |^2, \label{CResult}
\end{eqnarray}
where we have considered the direction of $d\bm{S}_x=d\vec{S}_x$, which is directed outwards from the metal surface.  In order to clarify the physical meaning of Eq. (\ref{CResult}), let us consider the surface between the metal and vacuum ($\alpha ^D = 0, \varepsilon ^D = 1$). In that case, Eq. (\ref{CResult}) can be reduced to the form below:

\begin{eqnarray}
\int d\bm{S}_x \frac{1}{8} \mathrm{Re} \left \{  \alpha ^{M*} \left ( \varepsilon ^M - 1 \right ) \right \}  | E_x^M(x_1^+) |^2 = \int d\bm{S}_x \left \{ \frac{1}{8}\alpha _R^M(\varepsilon _R^M -1) | E_x^M(x_1^+) |^2 + \frac{1}{8} \alpha _I^M \varepsilon _I^M  | E_x^M(x_1^+) |^2 \right \}. \label{CResult2}
\end{eqnarray}

The first term on the right hand side of Eq. (\ref{CResult2}) depends only on the real part of the polarizability and hence is the gradient force at the surface. On the other hand, the second term depends on the imaginary part of the polarizability and hence is the scattering force at the surface. Comparing Eq. (\ref{CResult2}) with Eq. (\ref{BulkForce}), we find that the OR voltage induced in the bulk region and that in the surface one are considered to be oppositely-oriented. Equation (\ref{CResult2}) is applied to the surface $\partial (S_1-S_2)$ at the boundary, whereas the integral (\ref{Surf-Force}) evaluated at the surface $\partial S_2$ tends to zero in the limit that $\epsilon \rightarrow 0$. Thus, the surface OR voltage induced in the divided part A is given as

\begin{eqnarray}
V_{x,\mathrm{DC,Surface}}^{(2),A} = \frac{1}{n^{(0)}q}\frac{1}{S_1} \int _{\partial (S_1-S_2)_x^{\mathrm{in}}}\left [ d\bm{S}_x  \frac{1}{8} \mathrm{Re} \left \{ \left ( \alpha ^{D}\frac{\varepsilon ^{M}}{\varepsilon ^{D}} + \alpha ^{M} \right )^* \left ( \frac{\varepsilon ^M}{\varepsilon ^D} - 1 \right ) \right \}  | E_x^{M,\mathrm{in}} |^2 \right ]_x, \label{VoltageSV} 
\end{eqnarray}
where $\partial S_x^{\mathrm{in}}$ denotes the surface just inside the metal surface and  $E_x^{M,\mathrm{in}}$ is the $x$ component of electric field just inside the metal surface to be integrated. The surface OR voltage at $x_2$ is also calculated in the same manner. Thus, the OR voltage across the structure is given by

\begin{eqnarray}
V_{x,\mathrm{DC}}^{(2)}= \frac{1}{n^{(0)}q}&&\Bigg [ \frac{\alpha _R}{4}\sum _{x=x_1, x_2} \bigg \{ \left \langle \left | \tilde{E}(x^-) \right |^2 \right \rangle _{S_2} - \left \langle \left | \tilde{E}(x^+) \right |^2 \right \rangle _{S_1} \bigg \}+\nonumber \\
&&\sum _{x=x_1, x_2} \frac{1}{S_1} \int _{\partial (S_1-S_2)_x^{\mathrm{in}}} d\bm{S}_x \Bigg [ \frac{1}{8} \mathrm{Re} \left \{ \left ( \alpha ^{D}\frac{\varepsilon ^{M}}{\varepsilon ^{D}} + \alpha ^{M} \right )^* \left ( \frac{\varepsilon ^M}{\varepsilon ^D} - 1 \right ) \right \}  | E_x^{M,\mathrm{in}}(x)|^2 \Bigg ]  \Bigg ]_x. \label{Formula_ApeA}
\end{eqnarray}
\end{widetext}

This is the formula for the OR voltage induced in a Lamellar metallic graing with the sharp edges. As well as Eq. (\ref{Formula_D}), the OR voltage depends on only the electric field intensities at $x_1$ and $x_2$. However, the derivation of this formula is rather mathematical and hence is difficult, at this stage, to obtain an intuitive understanding of the OR effect. We describe the interpretation of Eq. (\ref{Formula_D}) from the viewpoint of kinematics. Let us remind that light pressure is proportional to the density of light energy $W$, where $W = |\tilde{E} |^2/(4\pi)$.\cite{LandauKifshitzElectrodynamics}   Therefore, the factor in Eq. (\ref{Formula_D}),
\begin{eqnarray}
\left \langle \left | \tilde{E}(x_1) \right |^2 \right \rangle _S,
\end{eqnarray}
is proportional to the surface average of light pressure at $x=x_1$.  Based on the viewpoint of radiation pressure, the terms in Eq. (\ref{Formula_ApeA}) is interpreted as follows. The first and second term is the difference of light pressure across the surface. The third term is the light pressure just at the surface. Namely, the OR voltage originates from three contributions. As well as the discussion in the main text, it is concluded that the OR voltage generates where the cross section of the structure varies spatially.

\section{Electric field just inside the metal surface}
In Eq. (\ref{CResult}) electric field just inside the metal surface must be evaluated. In this appendix, we explain the procedure of the evaluation. We first numerically calculate electric field just at the surface at the position $x=x_1$: $E_x(x_1)$. On the other hand, the electric field just at the surface can be given by the space average of that in the dielectric and the metal bounding the surface:
\begin{eqnarray}
E_x(x_1) = \frac{E_x^D(x_1^-)+E_x^M(x_1^+)}{2}. \label{A1}
\end{eqnarray}
Combining Eq. (\ref{Conti-D}) and Eq. (\ref{A1}), we can obtain the electric field just inside the metal surface:
\begin{eqnarray}
E_x^{M}(x_1^+) = \frac{2}{1+\varepsilon ^M/\varepsilon ^D}E_x(x_1).
\end{eqnarray}
Electric field just inside the dielectric is also given by
\begin{eqnarray}
E_x^{D}(x_1^-) = \frac{2}{1+\varepsilon ^D/\varepsilon ^M}E_x(x_1).
\end{eqnarray}

\section{The OR polarization in a metallic Lamellar grating}

Following Eq.(\ref{Pol}), we can calculate the $x$ component of OR polarization induced in a Lamellar metallic grating shown in Fig. \ref{Fig4} (a):
\begin{eqnarray}
P_x \propto  q\frac{\alpha _R}{4}\int dS_x \left \{ \left | \tilde{E}(x_2) \right |^2 - \left | \tilde{E}(x_1) \right |^2 \right \}, \label{Px}
\end{eqnarray}
where the integral in (\ref{Px}) is taken over the interface between metal and dielectric. Comparing Eq. (\ref{Formula_D}) with Eq. (\ref{Px}), we find that the direction of the polarization induced in the structure,  in some cases such as resonance state, can be oppositely oriented to that of voltage across the structure. The surface average in Eq. (\ref{Formula_D}) is defined as
\begin{eqnarray}
 \left \langle \left | \tilde{E}(x) \right |^2 \right \rangle _S = \frac{1}{S_1}\int _{\partial S_1}dS_x \left | \tilde{E}(x) \right |^2, \label{Expression1}
\end{eqnarray}
where the surface integral is taken over the whole area of cross section at $x$. On the other hand, the surface integral in Eq. (\ref{Px}) is written as
\begin{eqnarray}
\int _{\partial (S_1-S_2)} dS_x \left | \tilde{E}(x) \right |^2, \label{Expression2}
\end{eqnarray}
where the surface integral is taken only over the surface between the metal and dielectric located at $x$. Comparing (\ref{Expression1}) with (\ref{Expression2}), we find that the integral in  (\ref{Expression1}) is nearly equal to (\ref{Expression2}) when the contribution of $S_2$ to the integral in (\ref{Expression2}) is sufficiently small, which corresponds to the limit that $S_2 \rightarrow 0$ or some resonance states where the electric fields at the surface $S_1$ are dominant compared with that of $S_2$.  Therefore, the OR voltage across the structure has, in some cases, opposite polarity to the OR polarization induced in the structure, which indicates that the second-order electric field induced by the OR effect, $\bm{E}_{\mathrm{DC}}^{(2)}$, is related to the depolarization field of the OR polarization.  Note that the direction of the OR polarization is not always oppositely oriented to the OR voltage.

\bibliography{Kurosawa_PRB_2014Ref.bib}

\end{document}